\documentclass[a4paper,11pt]{article}
\usepackage[top=3cm,bottom=3cm,left=3cm,right=3cm]{geometry}
\usepackage[utf8]{inputenc}

\usepackage{xcolor}
\usepackage{hyperref}
\usepackage{amsmath}
\usepackage{amssymb}
\usepackage{enumerate}
\usepackage{lipsum}
\usepackage{graphicx}
\usepackage{tabularx}
\usepackage{braket}
\usepackage{amsthm}
\usepackage{dirtytalk}
\usepackage{tikz}
\usepackage{commath}
\usepackage{qcircuit}
\usepackage{lineno}
\usepackage{authblk}
\begin{document}

\title{A Brief Review of Quantum Machine Learning \\ for Financial Services}

\author[,1]{Mina Doosti\thanks{mdoosti@ed.ac.uk}}
\author[1]{Petros Wallden}
\author[2]{Conor Brian Hamill}
\author[2]{Robert Hankache}
\author[3]{Oliver Thomson Brown}
\author[1]{Chris Heunen}

\affil[1]{\small \emph{School of Informatics, Quantum Software Lab, University of Edinburgh, United Kingdom}}
\affil[2]{\small \emph{Data Science \& Innovation, NatWest Group, London, United Kingdom}}
\affil[3]{\small \emph{EPCC, Quantum Software Lab, University of Edinburgh, United Kingdom}}

\date{}
\maketitle

\begin{abstract}
This review paper examines state-of-the-art algorithms and techniques in quantum machine learning with potential applications in finance. We discuss QML techniques in supervised learning tasks, such as Quantum Variational Classifiers, Quantum Kernel Estimation, and Quantum Neural Networks (QNNs), along with quantum generative AI techniques like Quantum Transformers and Quantum Graph Neural Networks (QGNNs). The financial applications considered include risk management, credit scoring, fraud detection, and stock price prediction. We also provide an overview of the challenges, potential, and limitations of QML, both in these specific areas and more broadly across the field. We hope that this can serve as a quick guide for data scientists, professionals in the financial sector, and enthusiasts in this area to understand why quantum computing and QML in particular could be interesting to explore in their field of expertise.
\end{abstract}

\section{Introduction}

Quantum computing, as a revolutionary and fundamentally different paradigm in computation, has significantly impacted diverse domains, including machine learning and data science. As machine learning continues to spearhead technological advancements in data science, the convergence of quantum computing and machine learning has garnered considerable interest, with the aspiration of transcending the boundaries of classical machine learning methodologies. The inception of Quantum Machine Learning (QML) dates back to the mid-1990s, initially exploring the concepts of quantum learning theory~\cite{bshouty1995learning}. However, it wasn't until approximately 18 years ago that the field gained significant attention following the seminal paper by Harrow, Hassidim, and Lloyd~\cite{harrow2009quantum}. This pivotal work laid the foundation for numerous studies in both supervised and unsupervised learning domains~\cite{havlivcek2019supervised,lloyd2018quantum,wiebe2012quantum,aimeur2007quantum,lloyd2014quantum,rebentrost2014quantum,schuld2014quest,cerezo2021variational}. Leveraging non-classical properties such as entanglement, superposition, and interference, quantum algorithms offer the potential for substantial speedups, sometimes exponential, over classical computing counterparts. While such speedups have been demonstrated for certain specific problems, achieving them in the realm of data science, and moreover for industry applications, remains an ongoing challenge with much active research in QML attempting to address it.

In this brief and concise review, instead of a broad examination of the field of quantum machine learning (as done in~\cite{cerezo2022challenges,biamonte2017quantum,abbas2021power}), we shift our focus to techniques and algorithms that parallel classical methods utilized in finance. We aim to highlight the potentials, promises, and challenges of applying QML within the financial sector. This review can also serve as a guide for financial professionals, both data scientists and management, offering an understanding of the latest advancements in QML relevant to their sector. Doing so, this review can aid decision-making processes, enhance the adoption of cutting-edge technologies, and lead to a deeper and hype-free appreciation of how QML can improve financial services.

Broadly categorized, QML encompasses three areas based on the nature of the algorithm and data: quantum algorithms on classical data, classical algorithms on quantum data and quantum algorithms on quantum data. Our review is centred on the first category, where classical data interfaces with quantum algorithms, as it aligns with the applications in finance, such as credit scoring, risk management, stock price prediction, and fraud detection. Within this scope, we explore quantum counterparts to Gradient Boosted Tree techniques for supervised learning, Generative AI models, and Graph Neural Networks, touching upon the most recent advancements in these topics.

First, we review some of the main and most widely used techniques from classical machine learning in financial applications. Then, we provide a brief, high-level overview of the promises and limitations of quantum machine learning, offering insights into the current landscape and prospects of this rapidly evolving field. Drawing insights from classical techniques, we select three categories of QML algorithms and review them with the desired applications in finance in mind. Finally, we conclude the review with some discussions.

\section{Applications of Classical Machine Learning in Finance}
In this section, we briefly review the various ways classical machine learning techniques are employed in financial services. These applications range from credit scoring and risk management to fraud detection, stock price prediction, and personalization through recommendation systems. Each of these areas benefits from the advanced capabilities of machine learning to improve accuracy, efficiency, and effectiveness. Below, we delve into specific techniques and their impacts in these key financial domains.

\begin{itemize}
    \item Credit scoring: Estimating the risk associated with lending to a particular individual or business is a key challenge in financial services. The measure of risk associated with selling loans typically informs the price offered to an individual or business and is a key component in ensuring that a financial institution operates safely and sustainably. 
    Traditionally, credit risk modelling has been undertaken in banks using linear algorithms like logistic regression \cite{mestiri2012credit}. However, the availability of large datasets and computing power has enabled the adoption of more modern machine learning and deep learning techniques. Studies comparing these modern techniques to traditional algorithms have found that deep learning outperforms logistic regression modelling, with gradient-boosted trees often identifying customer risk most accurately \cite{addo2018credit, shi2022machine}.
    Furthermore, graph convolutional networks have been shown to improve credit risk estimation by extracting features representing a company’s financial environment \cite{martinez2019graph}, indicating these algorithms can provide a more holistic view of a business to estimate credit risk.

    \item Risk management and compliance: Lenders are obliged to operate within an established risk appetite and meet compliance requirements. Estimating the portfolio risk is a key challenge for investment banks. Previous work has compared several different methods for estimating the value-at-risk in a stock portfolio, with the gradient-boosted algorithm Adaboost the most performant \cite{behera2023prediction}. Recently, the ability of Generative AI to create human-like text has been adopted by several banks to ensure their communications comply with regulatory frameworks \cite{bardeapplications}. As this technology matures, its usage is expected to become increasingly widespread, with the potential for substantially streamlining compliance processes. 

    \item Fraud detection: The automated detection of fraudulent transactions and behaviours allows lenders to reduce financial losses while protecting customers and building trust, without the cost and inaccuracy that manual analysis of transactions can entail. Popular algorithms that show strong performance in predicting fraud include support vector machines (SVMs) \cite{gyamfi2018bank}, boosted trees \cite{randhawa2018credit}, and artificial neural networks \cite{behera2015credit}. 

    \item Stock price prediction: Forecasting of future stock prices is a crucial but challenging task in financial services, informing investment strategies and risk management. Deep-learning-based techniques including the use of a Long Short-Term Memory (LSTM) network can capture complex relationships and outperform traditional forecasting methods like ARIMA \cite{yu2020stock}. Additionally, techniques like Generative Adversarial Networks (GANs) that generate new data, can be used to effectively classify and forecast time series \cite{vuletic2024fin}.

    \item Personalisation and recommendation systems: Machine learning has allowed different types of financial services to provide an increasingly personalised experience to customers, a vital aspect in an industry that consists of long-lasting and product-centric relationships between lenders and customers. Personalised contact marketing campaigns can be supported by the estimation of customer lifetime value (CLV), a metric of the total profit a lender may make from a customer over the course of their relationship. Machine learning allows more accurate predictions of CLV, enabling the marketing of relevant financial products to customers \cite{cowan2023modelling}. Recommendation of personal portfolios has been shown to be improved by the adoption of graph networks, in addition to a transformer-based neural network architecture to capture the relationship between financial products and how users behave over time \cite{li2024deep}.
\end{itemize}

\section{General Promises and Limitations of QML}
Before getting into the specific quantum machine techniques, let us first give a high-level summary of what can or cannot be expected from QML in general, by listing the most important promises and limitations.\footnote{Disclaimer: This is an opinionated list by the authors adopted from many different studies and discussions in the field, and it is by no means a comprehensive list. We refer to the following valuable QML reviews for similar arguments and more details~\cite {cerezo2022challenges,biamonte2017quantum,abbas2021power}.}\\

\noindent \textbf{Promises of Quantum Machine Learning:}
\begin{itemize}
    \item Efficient Training with Quantum-Aware Optimizers: QML offers the promise of efficient training through the development of quantum-aware optimizers that adapt to the unique characteristics of quantum computing, potentially outperforming classical optimization methods.

    \item Improved Generalisation: Recent studies suggest that QML models have the potential to achieve good generalisation even with a small amount of training data, offering the prospect of accurate predictions on previously unseen data.

    \item Quantum Advantage in Specific Tasks: QML holds the potential for achieving a quantum advantage in tasks like hidden parameter extraction from quantum data, generative modelling, and discovering quantum error correcting codes, particularly when dealing with quantum-mechanical processes. Also QML in some specific tasks can provide provable promises where there are no such counterparts in classical machine learning.

    \item Improved Accuracy: Quantum machine learning techniques offer the potential to achieve higher accuracy compared to classical models in various tasks such as classification, regression, and generative modelling, by leveraging quantum-enhanced feature spaces and complex data relationships. This is perhaps one of the most important advantages of using quantum models, even when they will not offer any speedups. These advancements can lead to more reliable predictions and better decision-making in fields.

    \item Quantum Advantage Beyond Speedup: Even though quantum algorithms are often been judged by whether or not they achieve a quantum speedup over a classical analogue, QML presents a unique opportunity for looking at different figures of merits where quantum computing can provide an advantage, for instance, privacy.
    
    \item Transition to Fault-Tolerant Quantum Computing Era: As quantum computers evolve into the fault-tolerant era, QML is expected to become even more useful, leveraging the capabilities of error-corrected quantum hardware to learn, infer, and make predictions directly from quantum data.
\end{itemize}

\noindent \textbf{Limitations of Quantum Machine Learning:}
\begin{itemize}

    \item Data Uploading on Quantum Memory: While QML models show promise with quantum data, the process of uploading the encoded classical data into quantum states on quantum registers and memories, namely QRAM, remains a significant challenge, especially when large datasets are involved.

    \item Challenges in Quantum Training Landscape: Many QML algorithms encounter challenges in training, such as local minima and Barren plateaus, which are analogous to the issue of vanishing gradients in classical ML. However, the additional complexity of the quantum landscape, coupled with the presence of hardware noise, amplifies these issues. Additionally, QML models tend to exploit the full space of quantum states in order to be more expressive, which often leads to the emergence of Barren plateaus.~\footnote{Quantum states are described in a vector space known as Hilbert space, which expands exponentially with system size. To fully utilize quantum systems' power, models tend to access all operations in this space, often referred to as expressibility. However, these highly expressive cases are the ones that often face vanishing gradient issues and training difficulties referred to as Barren plateaus. (See~\cite{mcclean2018barren,cerezo2021cost}}). Moreover, since techniques that have demonstrated effectiveness in classical ML have not yet been fully adapted and studied in the quantum setting, there is hope that future research will discover innovative approaches to mitigate these limitations.

    \item Impact of Quantum Noise: Hardware noise in quantum computations poses a critical challenge for QML in the pre-quantum-error-correction era, affecting all aspects of model performance and requiring strategies like error mitigation techniques or noise-resilient model design to address.

    \item Uncertainty Regarding Quantum Advantage: While exponential quantum advantage holds promise for specific tasks, there remains uncertainty regarding its realisation, especially in scenarios involving purely classical data, where achieving exponential advantages may not be feasible. As research in QML progresses, it becomes increasingly evident that there are instances where the likelihood of attaining a quantum advantage over classical algorithms is minimal or nonexistent~\cite{Huang2021InformationtheoreticBO,landman2022classically}. It is crucial to acknowledge that quantum advantage is not universally attainable or practical for all tasks. However, it is also important to note that Classical ML models often lack provable performance guarantees or even theoretical explainability of their performance, yet remain useful in countless practical scenarios. For this reason, perhaps, discarding QML algorithms solely due to the lack of exponential advantage might be too harsh of a verdict. It's worth noting that the criteria for demonstrating quantum computational supremacy, where rigorous proofs are needed, differ from those for demonstrating that a (quantum) model performs better than its classical competitors.~\footnote{We also refer to~\cite{Schuld2022IsQA} for more argument on why quantum advantage might not be the right figure-of-merit for QML.} 

\end{itemize}

\section{QML Algorithms for Supervised Learning Tasks}
In the realm of supervised learning tasks, where models learn from labelled data to make predictions or decisions, several QML approaches and architectures have been proposed in the literature, with some of them holding promising results for outperforming classical approaches. This section explores the application of quantum algorithms and techniques to address supervised learning tasks, with a focus on methods and techniques that are often used in data science and finance, such as Gradient Boosted Trees (GBT), or Quantum Neural Network (QNN).

\subsection{Quantum analogue for GBT?}
Gradient Boosted Trees (GBT) is a machine learning technique related to decision trees where combining multiple weak learners (typically shallow decision trees) creates a strong predictive model. GBT can be used for both classification and regression tasks, making them versatile in supervised learning. It is worth mentioning that the decision tree has been studied in~\cite{lu2014quantum} where it leverages quantum entropy for node splitting and quantum fidelity for data clustering while using Grover's search algorithm~\cite{grover1996fast} for searching over the tree to enhance the efficiency. However, the topic has not been further explored in quantum machine learning. As a result, due to the lack of an exact quantum equivalent technique, in this section, we focus on relevant techniques to the applications for which GBT is used.

While GBT does not have an exact quantum equivalent to this date, QML offers several alternative methods for enhancing supervised learning. The main parallel lies in enhancing supervised learning tasks through quantum-enhanced feature spaces. These quantum algorithms aim to expand feature space representation in higher dimensions, potentially capturing complex data relationships, and thereby improving task accuracy, which is the focus of the next section.

\subsection{Supervised learning with quantum-enhanced feature spaces}
This work~\cite{havlivcek2019supervised} introduces two quantum algorithms for near-term quantum devices: the Quantum Variational Classifier, resembling conventional SVMs, and Quantum Kernel Estimation, for optimizing a classical SVM on a quantum computer. Both utilize the quantum state space as a feature space and offer advantages hard to simulate classically, following the concept of quantum SVMs introduced in~\cite{rebentrost2014quantum}. The study also explores applications for noisy intermediate-scale quantum computers in machine learning. The Quantum Variational Classifier operates in four steps: mapping classical data to a quantum state, applying a short-depth quantum circuit, performing binary measurements, and using a decision rule based on empirical distribution. Its experimental results show a classification success rate close to $100\%$ for circuit depths~\footnote{Circuit depth in quantum computing refers to the number of sequential layers of quantum gates (over one or several qubits) required to implement a quantum algorithm or circuit. It is related to the time complexity of the circuit and is a crucial factor in assessing the computational resources needed for quantum computations. A deeper circuit often implies a longer execution time, during which the qubits might decohere and lose their useful quantum information, hence long-depth quantum circuits are more challenging to implement on quantum hardware.} greater than 1, with optimal performance at depth 4, even considering decoherence effects. The classifier is trained on three datasets per depth and performs 20 classifications per trained set, using 20,000 shots~\footnote{In quantum computing experiments, "shots" refer to the number of times a quantum circuit is executed at the end of which a fixed quantum measurement is repeated. Each shot represents an independent execution or measurement of the quantum system, providing statistical information about the outcomes of the experiment. By averaging the results over multiple shots, one can obtain an estimate of probabilities or expectation values associated with quantum states or operations, which is often necessary given the probabilistic nature of quantum mechanics.} for running classification experiments compared to 2,000 for training. The Quantum Kernel Estimation algorithm estimates the SVM kernel on a quantum computer for both training and classification phases. The quantum computer initially estimates the kernel for all pairs of training samples. In the classification phase, the quantum computer is used to estimate the kernel for a new data point using the support vectors obtained from the optimization, providing sufficient information to construct the complete SVM classifier. The estimation of inner products for the kernel involves direct estimation from transition amplitudes, utilizing feature map circuits. The transition probability is estimated by measuring the final state in the usual computational basis multiple times. It achieves high classification success, with two test sets reaching $100\%$ and a third averaging $94.75\%$ success~\footnote{For more details, see FIG. S7 (page 22) in ~\cite{havlivcek2019supervised}. Accuracy and experimental data are provided, however, no comparison with classical is given, most probably since the experimental accuracy was very high.}. This method's effectiveness is underlined by its ability to maintain the positive semidefiniteness of the kernel despite sampling errors.

Further studies, such as~\cite{schuld2019quantum} and~\cite{schuld2020circuit}, delve deeper into these quantum approaches. Particularly, \cite{schuld2020circuit} benchmarks the circuit-centric quantum classifier against classical techniques, testing it on datasets like CANCER, SONAR, WINE, SEMEION, and MNIST256. Despite fewer parameters, the quantum classifier often outperforms or matches classical models like MLPshal and MLPdeep, although it shows overfitting in some cases, suggesting the need for improved regularisation techniques.

\subsubsection{Provable QML algorithms in this area}
Most notable recent developments in this area (since 2023) include~\cite{jager2023universal}, demonstrating the quantum advantage of variational quantum classifiers and quantum kernel SVMs in solving certain hard problems in complexity theory~\footnote{To be specific, the problem studied in this paper is PROMISEBQP-complete.}. This implies potential efficiency in Bounded-Error Quantum Polynomial-Time decision problems. Another significant contribution is~\cite{gentinetta2024complexity}, which highlights the efficiency of quantum SVMs. Using quantum circuits to define kernel functions, it achieves a provable exponential speedup over classical algorithms for specific datasets. The complexity of solving the dual formulation is estimated at $O(M^4.67/\epsilon^2)$ quantum circuit evaluations, and empirical analysis suggests addressing the kernelized primal problem in $O(min{M^2/\epsilon^6, 1/\epsilon^{10}})$ evaluations, where $M$ denotes the size of the data set, and $\epsilon$ the solution accuracy compared to the ideal result from exact expectation values, which is only obtainable in theory. This work also presents a variational approximation to quantum SVMs, showing improved scaling in heuristic training.

\subsection{Quantum Neural Networks}
Many QML models rely fundamentally on Parameterized Quantum Circuits (PQCs) as a vital component. These circuits consist of a series of parameterised unitary gates acting on quantum states which encode the classical data~\cite{cerezo2022challenges}. The parameterized circuits include quantum gates that contain free parameters $(\theta)$ that are adjusted through training to address specific problems. PQCs share a conceptual quantum analogue of neural networks, and it has been shown that classical feedforward neural networks can be formally embedded into PQCs~\cite{wan2017quantum}. In the QML literature, certain types of PQCs are often referred to as QNNs, and sometimes QNNs are considered as a subclass of Variational Quantum Algorithms (VQA)~\cite{abbas2021power}. In practical terms, the term QNN is utilized whenever a PQC is applied in a data science context (mostly when the data is classical and the algorithm is quantum). Specifically related to supervised classification tasks, a QNN aims to map states from different classes to distinguishable regions within the Hilbert space.

QNNs can be realised in different architectures. As a simple categorisation, one can look at three different examples where in each layer the number of qubits in the model is increased, preserved or decreased (like dissipative QNNs or convolutional QNNs). Dissipative QNNs generalize the classical feedforward network~\cite{cerezo2022challenges} as mentioned. In a standard QNN architecture, the classical data is encoded into quantum states and sent to a QNN where the quantum circuit is applied and at the end all of the qubits are measured. This is an example of a QNN architecture where the number of qubits is preserved. Convolutional QNN, studied in~\cite{cong2019quantum}, and demonstrated for having good classification performance, is an example of a QNN architecture where the number of qubits is reduced since qubits are measured and discarded in each layer to reduce the dimension of the data while preserving its relevant features. Also, the trainability and capacities of QNNs have been studied in~\cite{abbas2021power} where implementation on real hardware has also been demonstrated. In general, QNNs and their different architectures are one of the main and active areas of research in QML~\cite{farhi2018classification,cerezo2022challenges,bausch2020recurrent}.

\subsection{QML techniques related to credit scoring and risk management}
In terms of applications for finance, recent studies have utilized quantum machine learning (QML) for credit scoring and financial risk measurement. First, we introduce the paper \cite{leclerc2023financial} where a quantum-enhanced machine learning method for predicting credit rating has been proposed~\footnote{Here in this work, ``quantum-enhanced" specifically points that (classical) classifier is trained with a quantum adiabatic algorithm. The method referred to as QBoost developed in~\cite{neven2008training}. However the term ``quantum-enhanced" is used in different meanings in the literature, including hybrid (quantum-classical) algorithms, and quantum-inspired classical algorithms.}. Implemented on a neutral atom quantum processor with up to 60 qubits, this model shows promising results, with competitive performance, improved interpretability, and training times comparable to state-of-the-art random forest models. In \cite{thakkar2023improved}, authors explore quantum machine learning for enhancing financial forecasting. They incorporate classical and quantum Determinantal Point Processes in Random Forest models, achieving nearly $6\%$ improvement in precision, compared to the classical random forest. Additionally, they design quantum neural network architectures with orthogonal and compound layers for credit risk assessment. These models match classical performance but with significantly fewer parameters, demonstrating the efficiency of quantum methods in machine learning.
Lastly, \cite{wilkens2023quantum} examines the broader potential of quantum computing in financial risk management, focusing on Value-at-Risk and Potential Future Exposure. The study acknowledges the feasibility of conceptual solutions and small-scale circuits but also highlights challenges in real-life applications, such as the need for increased hardware capacity (more qubits) and quantum noise mitigation. In this area of applications, one of the first works is by~\cite{milne2017optimal}, where the authors demonstrated translating these challenges into a quadratic unconstrained binary optimization (QUBO) problem, solvable by quantum annealers. This approach was tested on a quantum simulator, indicating the potential of quantum annealers in optimizing credit analysis features (further detailed in \cite{orus2019quantum}).

\section{Quantum Computing for Generative AI}
Generative AI or deep Generative Learning Models (GLMs) have revolutionized the landscape of artificial intelligence and have been successfully applied to various tasks~\cite{goodfellow2014generative,blei2017variational}. With classical GLMs approaching the computational limitations according to Moore's law~\cite{thompson2007computational}, and the unique abilities of quantum computers in handling large dimensionalities and producing complicated probability distributions, Quantum Generative AI (QGenAI) represent a promising field for enhancing computational efficiency and overcoming these barriers and paving the way for the next generation of artificial intelligence. Several research directions have been proposed in this domain which we can broadly categorise as Quantum circuit Born machine (QCBM)~\cite{benedetti2019generative}, quantum Boltzmann machine (QBM)~\cite{amin2018quantum}, quantum generative adversarial network (QGAN)~\cite{lloyd2018quantum}, and Quantum Transformers~\cite{kerenidis2024quantum,khatri2024quixer}. Given that classical transformers are state-of-the-art technology for many applications of interest, in this section, we mainly focus on their quantum analogue and we will only briefly mention other techniques.

\subsection{Quantum Transformers}
The key component of the transformer, is the attention mechanism, as introduced by~\cite{vaswani2017attention,dosovitskiy2020image}. The process involves transforming input data (of dimension $n \times d$), denoted as $X \in \mathbb{R}^{n \times d}$, into n patches with dimensions of $d$. The trainable weight matrix from the initial fully connected layer is denoted as $V$. The core of the attention mechanism is represented by the attention coefficients which assign weights to each patch concerning every other patch (through the equation $A{ij} = x^T_i \textbf{W} x_j$, where $\textbf{W}$ is the new trainable weight matrix). A quantum transformer also focuses on the attention mechanism. In a recent work~\cite{kerenidis2024quantum}, three types of quantum transformers have been introduced, building upon previous related works~\cite{cha2021attention,di2022dawn,li2022quantum} and applied to visual tasks for benchmarking: Orthogonal Patch-wise Neural Network, Quantum Orthogonal Transformer and Qunatum Compound Transformer. The first one implements a trivial attention mechanism where each patch pays attention only to itself. The second one is designed to be a close quantum analogue of the classical approach and capture both a linear fully connected layer, and the attention matrix to capture the interaction between patches. The third one, distinct from the other two, harnesses the power of quantum superposition in loading data and is inspired by the MLPMixer architecture~\cite{tolstikhin2021mlp}. The compound transformer introduces a unified operation that can replace the former ones and facilitates information exchange between patches without relying on convolution or attention mechanisms. Now let us have a brief overview of how each of these quantum transformers work.\\

\noindent \textbf{The Orthogonal Patch-wise Neural Network} simplifies the attention mechanism, where as mentioned each patch only pays attention to itself. i.e. each input patch is multiplied by the same trainable matrix $V$, with one circuit per patch. The quantum circuit employs a quantum orthogonal layer to perform this multiplication. The output of each circuit is a quantum state encoding $V x_i$, a vector retrieved through tomography~\footnote{tomography refers to the process of measuring multiple copies of a quantum state to retrieve its classical description.}. Interestingly, the tomography procedure avoids exponential complexity as it deals with states of linear size in this case. A data
loader with $N = d$ qubits is required, and computational complexity is given as $O(\log(d))$, with trainable parameters $O(d \log(d))$ and fixed parameters requiring $d - 1$. This quantum transformer is designed to achieve attention without the intricate computations associated with classical attention mechanisms.

\noindent \textbf{The Quantum Orthogonal Transformer} operates by calculating each attention coefficient $A_{ij}$ by loading $x_j$ into the circuit with a vector loader, followed by a trainable quantum orthogonal layer $\textbf{W}$, resulting in the vector $\textbf{W} x_j$. Subsequently, an inverse data loader of $x_i$ is applied, creating a state where the probability of measuring 1 on the first qubit is exactly the desired attention $|x^T_i \textbf{W} x_j|^2 = A_{ij}^2$. The positive coefficients for $A$, are learned during training. Then a post-processing with softmax is applied to obtain each coefficient, and these components are classically combined. The computational complexity of this quantum circuit is similar to the Orthogonal Patch-wise Neural Network.

\noindent \textbf{The Quantum Compound Transformer} represents a shift towards a more native quantum approach. In this transformer, all patches are loaded as a quantum state with the same weight, in superposition, and then processed by an orthogonal layer. This orthogonal layer simultaneously extracts features from each patch and re-weights them. The output is computed as a weighted sum of the features extracted from all patches. The process is thus simplified by using one operation instead of calculating separate weight matrices. The quantum circuit which executes one attention layer of this transformer employs a matrix data loader for the input matrix $X$ and a quantum orthogonal layer for $V$, applied on both registers. In summary, the Quantum Compound Transformer handles feature extraction and weighting in a single operation and provides a computational complexity of $O(\log(n) + n \log(d) + \log(n + d))$.

We refer to Table 2 in~\cite{kerenidis2024quantum}, for a comprehensive comparison between these quantum transformers and state-of-the-art classical approach. The authors also provide simulations and experiments to benchmark their proposed quantum methods for medical image classification tasks. The experiments were performed on the MedMNIST dataset, which comprises 12 preprocessed, two-dimensional medical image datasets. Besides the three quantum transformers, two baseline methods Vision Transformer and Orthogonal Fully-Connected Neural Network (OrthoFNN) were also included for comparison. The results show that the Vision Transformer, Quantum Orthogonal Transformer, and Quantum Compound Transformer outperformed Orthogonal Fully-Connected and Orthogonal Patch-wise neural networks across all 12 MedMNIST tasks. But more importantly, quantum architectures achieved comparable accuracy to classical benchmarks while utilizing fewer trainable parameters. The resource analysis for simulated quantum circuits revealed promising results for the Compound Transformer which required 80 trainable parameters compared to 512 $(2d^2, d=16)$ for the Classical Vision Transformer. Circuit depth and the number of distinct circuits were provided for each quantum architecture. We also refer to Table 5 in~\cite{kerenidis2024quantum}, for the exact number of qubits, circuit depth and parameters for each. The work also includes experiments on IBM Quantum devices that demonstrated competitive accuracy levels for quantum transformers compared to classical benchmarks (however slightly lower). Hence this work shows an important potential for quantum transformers compared to their classical counterparts in terms of improving accuracy and lowering the number of required training parameters.

Additionally, a recent preprint~\cite{guo2024quantum} investigates the integration of fault-tolerant quantum computing into transformer architectures by employing techniques from the quantum signal processing framework and quantum singular value transformation to construct efficient quantum subroutines for key transformer components, and claims to achieve a quadratic advantage under specific parameter regimes.

Finally, very recently, a new architecture for quantum transformed, namely \emph{Quixer} has been introduced in~\cite{khatri2024quixer}, which utilises the Linear Combination of Unitaries~\cite{childs2012hamiltonian} and Quantum Singular Value Transform primitives as building blocks. The model is also applied to a small practical language modelling task, and resource estimation has been provided~\footnote{Disclaimer: We do not compare the architecture of~\cite{khatri2024quixer} with the one presented in more detail from the work of~\cite{kerenidis2024quantum} or able to provide much more details, primarily because the new paper was accessible quite recently after the main body of this review had been completed.}.

Another remarkable and rather different study related to quantum transformers is presented in~\cite{anschuetz2023interpretable} and investigates the comparative expressive power of quantum contextual recurrent neural networks (CRNNs) and classical linear recurrent neural networks (LRNNs) for sequence-to-sequence learning tasks. The authors demonstrate that CRNNs, incorporating quantum contextuality—a fundamental feature of quantum systems—exhibit a quadratic memory separation advantage over classical models, and indicate a practical advantage for generalisation of smaller models. Intuitively, quantum contextuality is similar to linguistic contextuality where the meaning of a word depends on the sentence \textit{context} determined by the other words. Taking inspiration from this similarity, the authors evaluate their quantum model against state-of-the-art classical models on a standard Spanish-to-English translation task. Comparative performance analysis with LRNN, GRU RNN, classical transformer, and Gaussian model illustrates the quantum advantage, showing better performance across all model sizes. This work highlights intriguing and less explored directions to explore the potential advantage of QGenAI algorithms, by diving into more foundational quantum properties.

\subsection{Other QGenAI techniques}
\noindent \textbf{Quantum Circuit Born Machine (QCBM)} introduced by Benedetti et al.~\cite{benedetti2019generative}, serves as the quantum counterpart to classical Stochastic Neural Networks (SNN). In QCBM, randomness originates from intrinsic quantum mechanical properties rather than being sampled after each layer. The main idea is to use a Quantum Neural Network (QNN) to generate a tunable discrete probability distribution that approximates a target distribution. This is achieved by leveraging a Parametrized Quantum Circuit (PQC) to manipulate an initial quantum state and generate the desired distribution. The performance of QCBM heavily relies on the topology of the entangling layers in PQCs, with high performance achieved when the topology matches the hardware architectures. QCBM has been applied to various generative learning tasks~\cite{tian2023recent}. Applications of QCBM extend to finance, where it has been used for learning empirical financial data distributions~\cite{alcazar2020classical,kondratyev2021non}, generating synthetic financial data~\cite{coyle2021quantum,kondratyev2021non}, and learning joint distributions~\cite{coyle2021quantum,alcazar2020classical}, showing heuristic advantages over classical models, particularly as problem scale increases~\cite{pistoia2021quantum}.

\noindent \textbf{Quantum Boltzmann  Machine (QBM)} is a quantum version of the classical Boltzmann machine (BM), proposed by Amin et al.~\cite{amin2018quantum}. It leverages quantum devices to prepare the Boltzmann distribution, which estimates discrete target distributions. In QBM, qubits replace the units in BMs, and the energy term in classical BMs' Hamiltonian is replaced by a quantum Hamiltonian. For instance, a transverse-field Ising model is commonly used, with the Hamiltonian formulated to include trainable parameters~\cite{amin2018quantum,tian2023recent}. QBM aims to minimize the negative log-likelihood of the target distribution by updating its parameters through optimization methods. An important aspect of QBM is its training process, which involves calculating gradients using both positive and negative phases. The positive phase refers to the Boltzmann average, while the negative phase sampling is NP-hard due to the exponential complexity of exact diagonalisation~\cite{barahona1982computational}. Several approximation methods, such as quantum Monte Carlo techniques, have been proposed to address this challenge~\cite{amin2018quantum,yung2012quantum,anschuetz2019realizing,xiao2020quantum}. QBM has been applied to simulate probability distributions, achieve better trainability than classical Boltzmann machines, and has been applied to many areas including language processing~\cite{wiebe2019quantum}, and finance~\cite{crawford2016reinforcement,alcazar2020classical,orus2019quantum}. QBMs are promising techniques in QGenAI and are a subject of ongoing active research in this domain.

\noindent \textbf{Quantum Generative Adversarial Network (QGAN)} is a novel approach in QGenAI initially conceptualized by Lloyd and Weedbrook~\cite{lloyd2018quantum}. QGANs follow the framework of classical GANs, employing a generator and a discriminator to engage in a two-player minimax game. However, QGANs differ in how these components are constructed, often utilizing QNNs instead of classical deep neural networks. QGANs can estimate both discrete and continuous distributions, offering potential computational advantages over classical GANs due to their utilisation of quantum resources~\cite{lloyd2018quantum,dallaire2018quantum,Zoufal2019QuantumGA}. Different types of QGANs have been explored for various tasks, including quantum chemistry calculations~\cite{Li2021InvitedDD}, image generation~\cite{Huang2020ExperimentalQG}, and finance~\cite{Assouel2021AQG,Pan2022ApplicationOQ}.

\section{QML Algorithms for Graph-related Problems}
In finance, Graph Data Science methods, encompassing topological analysis and algorithms for centrality, similarity, and anomaly detection, provide crucial insights into complex financial networks. Graph Neural Networks (GNNs) further extend these insights by leveraging graph data for supervised learning tasks such as fraud detection and stock price prediction. Within the quantum realm, efforts have emerged to explore the capabilities of quantum computing and quantum machine learning, notably through Quantum Graph Neural Networks (QGNNs), a subset of Quantum Neural Networks (QNNs) which has been discussed earlier. The first work in this area was by Verdon et al.~\cite{verdon2019quantum} introducing the QGNN algorithm as a quantum graph classification model, aiming to harness QML techniques for graph-structured data and derive node representations in a quantum Hilbert space. The algorithm encodes the graph structure in a quadratic Hamiltonian and employs a parameterized quantum circuit, along with standard QNN techniques, to extract relevant graph information tailored to the application of interest. One notable application proposed in the paper involves employing the Quantum Graph Convolutional Network (QGCNN) for graph isomorphism classification, which is also used as a benchmark for evaluating the representation power of classical graph neural networks~\cite{xu2018powerful}. The implementation of QGCNN involves utilizing the QGCNN ansatz with single-qubit precision encoding, applied in parallel to the structures of two given graphs (G1 and G2). By sampling eigenvalues of the relevant coupling Hamiltonian on both graphs and conducting standard basis measurements, a set of `energies' of this Hamiltonian is obtained for classification purposes. 
The dataset used for training comprises graphs sampled uniformly at random from a specific distribution for a fixed number of nodes. The training process involves optimizing the networks using a Nelder-Mead optimization algorithm. Despite the relatively small sample size, the experiments demonstrate promising accuracy for classification, underscoring the potential efficacy of QGCNNs.

Follow-up studies include~\cite{mernyei2022equivariant}, addressing certain limitations in existing Graph Neural Networks (GNNs). This paper introduces the Equivalent Quantum Graph Circuit (EQGC). Their approach involves capturing permutation-invariant topologies of input graphs (albeit constrained by linearly scaling qubit requirements) limiting its applicability to small-scale synthetic datasets. The next important work is the Graph Quantum Neural Tangent Kernel (GraphQNTK)~\cite{tang2022graphqntk} that combines ideas from GNN and Graph Kernels (GK)~\cite{haussler1999convolution}. The work introduces a hybrid quantum-classical method where a quantum algorithm estimates the neural tangent kernel of the underlying classical graph neural network, effectively emulating an infinite-width GNN. They have also introduced a quantum attention mechanism to better capture the semantic similarity information of graph nodes into the model. This work stands out compared to the previous ones due to its capability to handle more realistic graph data, in terms of size\footnote{An open-source repository of their code can be found at \hyperlink{https://github.com/abel1231/graphQNTK}{https://github.com/abel1231/graphQNTK}.}. The authors compare their model with state-of-the-art GKs and GNNs such as WL kernel, AWL, RetGK, GNTK, WWL, GCN, PatchySAN, GCKN, GraphSAGE, and GIN (we refer to Table 2 in~\cite{tang2022graphqntk} for the comparison). 

Common challenges in QGNN are the absence of general methods for mapping data from Euclidean space into a quantum Hilbert space, as well as the fact that they typically accommodate low-dimensional data. These two challenges have been addressed in a recent pre-print by Ai et al.~\cite{ai2022towards} where they propose another hybrid quantum-classical algorithm namely Egograph-based Quantum Graph Neural Network (egoQGNN) for graph classification. Compared to the previous works, egoQGNN uses a hierarchical architecture, and a proposed decompositional processing strategy to capture information within k-hop neighbours of a node, and as a result claims to better handle real-world datasets, in addition to coming together with the proof of the classification capability and the first theoretical framework to describe the data mapping. This work also provides simulations and heuristic performance and comparison to classical counterparts.

%%%%%%%%%%%%%%%%%%%%%%%%%%%%%%%%%%%%%%%%%%%%%%%%
\section{Conclusion and Discussions}
This short review paper outlines QML algorithms with promising potential applications in finance. We explored quantum enhancements to classical machine learning techniques, focusing on supervised learning tasks, Generative AI models, and Graph Neural Networks. Our analysis highlights the potential of QML in tasks such as credit scoring, risk management, stock price predictions, and fraud detection within the financial sector. Upon reviewing our findings, it's evident that certain areas, algorithms, and techniques within QML offer greater promise and utility for finance, both in the near and long term.

In the near term, Quantum Variational Classifier and Quantum Kernel Estimation algorithms emerge as realistic options for application on Noisy Intermediate-Scale Quantum (NISQ) devices, showing potential for improvements in tasks like credit scoring and risk management. While some existing algorithms demonstrate high classification success rates, further experimentation on real hardware is necessary to validate their efficacy compared to classical methods. Even with the lack of provable exponential quantum advantage in this area, given their hybrid nature and the fact that these algorithms are not as expensive and resource-intensive as the fault-tolerant ones, adoption of these techniques could prove beneficial, particularly if they offer significant improvements in precision or other metrics. 

Looking ahead to the long term, Quantum Neural Networks present compelling opportunities for long-term advancements in financial analysis, especially given that they appear to be relevant in all three general areas discussed in this paper (Supervised learning, GenAI, and Graph-related problems). Given their ability to capture complex data relationships and simplify feature extraction, architectures like the Quantum Compound Transformer show promise for enhancing models used in financial forecasting and decision-making. By leveraging these unique features of quantum and combining them with techniques from the quantum signal processing framework, these approaches aim to overcome the computational limitations of current and future classical ML. However, they should be perhaps looked at as a longer-term investment due to the lack of sufficient research and experiments to demonstrate their real-life applicability which is currently out of reach of current and near-term quantum devices.

Moreover, Quantum Graph Neural Networks (QGNNs) hold promise for innovation in finance, particularly in tasks such as fraud detection and stock price prediction. Ongoing research addressing challenges like data mapping and scalability suggests significant potential for long-term revolutionizing of graph-related problems in finance. However, at the moment it is unclear whether or not the real-life application in this area can be achieved in the near term.

Despite the opportunities presented by QML, it's crucial to acknowledge the challenges and limitations, such as efficient data uploading and training challenges in the quantum landscape.

Looking forward, the future of QML holds promise for transformative advancements in machine learning and data science. By addressing existing challenges and fully harnessing quantum technologies, we can unlock new frontiers in computational science and drive innovation across diverse industries. Additionally, the study of quantum algorithms serves as inspiration for creating novel algorithms in classical machine learning, including the rise of perhaps new generations of quantum-inspired algorithms, highlighting the value of QML research from different application perspectives, irrespective of the technological challenges posed by quantum computing hardware.\\

\noindent\textbf{Acknowledgements:}
The authors acknowledge the support received from the strategic partnership between NatWest Group and University of Edinburgh, particularly Raad Khraishi and Aidan McFadden who provided insights on applicability of machine learning techniques in practice within financial services, and Paul Carter, Phoebe Choi and Kostas Kavoussanakis who managed this engagement. 

%%%%%%%%%%%%%%%%%%%%%%%%%%%%%%%%%%%%%%%%%%%%%%%%%%
\bibliographystyle{unsrt}
\bibliography{refs}
\end{document}